\documentclass[%
 reprint,
 amsmath,amssymb,
 aps,
]{revtex4-2}
\usepackage{subcaption}
\usepackage{amsmath}
\usepackage{graphicx}
\usepackage{dcolumn}
\usepackage{bm}

\usepackage{color}

\def\bu{\boldsymbol{u}}
\def\bs{\boldsymbol{s}}
\def\bv{\boldsymbol{v}}
\def\bf{\boldsymbol{f}}
\def\bff{\boldsymbol{f}}

\def\pvt{ \boldsymbol{\mathcal{P}}}

\newcommand{\ksi}{\xi}
\newcommand{\vksi}{\boldsymbol{\ksi}}
\newcommand{\herm}{\Psi}
\begin{document}

\preprint{APS/123-QED}

\title{Uncertainty quantification of time-average quantities of chaotic systems using sensitivity-enhanced polynomial chaos expansion}

\author{Kyriakos D. Kantarakias}
 \email{k.kantarakias@imperial.ac.uk}
\author{George Papadakis}%
 \email{g.papadakis@imperial.ac.uk}
\affiliation{ Department of Aeronautics, Imperial College London}%

\date{\today}

\begin{abstract}
We consider the effect of multiple stochastic parameters on the time-average quantities of chaotic systems.  We employ the recently proposed \cite{Kantarakias_Papadakis_2023} 
sensitivity-enhanced generalized polynomial chaos expansion,  se-gPC, to compute efficiently this effect. se-gPC is an extension of gPC expansion, enriched with the sensitivity of the time-averaged quantities with respect to the stochastic variables. To compute these sensitivities, the adjoint of the shadowing operator is derived in the frequency domain. Coupling the adjoint operator with gPC provides an efficient uncertainty quantification (UQ) algorithm which, in its simplest form, has computational cost that is independent of the number of random variables. The method is applied to the Kuramoto-Sivashinsky equation and is found to produce results that match very well with Monte-Carlo simulations. The efficiency of the proposed method significantly outperforms sparse-grid approaches, like Smolyak Quadrature. These properties make the method suitable for application to other  dynamical systems with many stochastic parameters.
\end{abstract}

\maketitle

\section{\label{sec:1}INTRODUCTION}
The performance of real world systems is significantly affected by the uncertainty of the parameters that define these systems. Large research effort has focused on the quantification of the effect of such stochastic variations to a Quantity of Interest (QoI), usually a time-averaged quantity. This field of research is commonly known as Uncertainty Quantification (UQ) and efficient UQ methods have been developed for static and dynamic problems \cite{UQ_Nature,GhanemSpanos,KnioBook, CH2019207}.  

In the standard generalized polynomial chaos (gPC) method, originally proposed in \cite{GhanemSpanos,KarniadakisOriginal}, an orthonormal polynomial base that spans the stochastic space is used for the spectral representation of uncertain quantities. The spectral coefficients are computed with Galerkin projection, allowing for the efficient evaluation of the statistics of the QoI. However, the cost of gPC scales as $\sim m^p$, where $m$ is the number of stochastic parameters and $p$ the polynomial order of the expansion. This exponential growth is known as the 'curse of dimensionality' \cite{KnioBook}. Various approaches have been proposed to mitigate the rapid growth of the computational cost, such as sparse grid approaches like Smolyak grids \cite{Smolyak_1963}, or adaptive methods that build a sparse polynomial chaos expansion (PCE) basis using least-angle regression \cite{BLATMAN20112345}.  gPC methods have been successful in predicting the statistics of the QoI in many applications, such as fluid dynamics, mechanics,  space, medicine, see for example \cite{GhanemSpanos, CHASSAING2012394,SCHIAVAZZI2017196,CH2019207,JONES20131860}. Applications of gPC to chaotic systems have also appeared in the literature  \cite{CHE2018208,UQChaos2,ChaosUQ2,PreKant,a13040090}.

Another method for the computation of the spectral coefficients is based on the least-squares approach \cite{KnioBook,BLATMAN2008518}. To reduce the computational coast, the method is usually coupled with efficient multidimensional sampling techniques; for an overview of the different sampling algorithms, see \cite{HADIGOL2018382}. Recently, new sampling approaches that incorporate real-world data into the computation of the gPC coefficients have been introduced \cite{OLADYSHKIN2012179,TORRE2019601,AHLFELD20161}. 
However the cost still grows exponentially with the number of uncertain parameters, making such methods difficult to apply in systems with a large $m$.

In this paper we use the sensitivity-enhanced gPC, or se-gPC, a least squares approach for the computation of the spectral coefficients that is augmented with the sensitivity of the QoI with respect to the uncertain parameters, see \cite{Kantarakias_Papadakis_2023} for a detailed description of the method and a review of previous works in this area. When all the sensitivities are computed efficiently in a single step with the adjoint method, the computational cost of se-gPC is reduced by a factor $m$, i.e.\ it scales as $\sim m^{p-1}$, and the method becomes increasingly useful as the number of stochastic inputs grows. Efficient sampling algorithms can by employed to reduce the number of required evaluations, see \cite{Kantarakias_Papadakis_2023} for comparison between two such algorithms. For the special case of first order spectral representation, i.e.\ for $p=1$, the spectral coefficients can be estimated with a single direct and a single adjoint evaluation, regardless of the number of stochastic inputs. 

The se-gPC is efficient because the sensitivities of the QoI with respect to all stochastic inputs at one sampling point can be estimated using a single adjoint evaluation. Adjoint methods have long been successfully used to estimate derivatives (sensitivities) for stationary or non-chaotic systems in aerodynamics \cite{adjointjameson}, structural optimization \cite{DEMS1984527}, chemical kinetic systems \cite{SANDU20035083} among many others. However, when the underlying system is chaotic, a small variation in an input parameter causes large deviation in the trajectory of the system in phase space (with respect to the reference trajectory); this is popularly known as 'butterfly effect'. Under these circumstances, standard sensitivity analysis tools (such as adjoint) fail to produce physically meaningful results \cite{Eyink_2004,Lea1}. Mathematically, the deviation between the two trajectories is due to the presence of one or more positive Lyapunov exponents. To address this problem, the least-squares shadowing method (LSS) and its variants were proposed in a series of papers \cite{WANG2014210,WangLSS,BLONIGAN201416}. This method is relies on the shadowing lemma \cite{Pilyugin1999ShadowingSystems,Holmes2012TurbulenceSymmetry} and provides a systematic and rigorous approach for the computation of sensitivities of time-average quantities of  chaotic systems. 

Assume a reference trajectory of a dynamical system evaluated for a parameter value $s$, $\boldsymbol{u}(t;s)$. Shadowing  methods aim to compute another trajectory at $s+\delta s$,  $\boldsymbol{u}(\tau;s+\delta s)$ that 'shadows', or stays close to, the reference trajectory for the time frame $T$ of the analysis. The sensitivity problem is reformulated as a minimization problem between the reference and the shadowing trajectories. This formulation regularizes the problem, the two trajectories remain close to each other, and can be used to compute accurate sensitivities for a long $T$. Note that  the two trajectories start from different initial conditions,  but for ergodic system this does not affect the time-average QoI. Also, the time variable in the shadowing trajectory is  not be the same as $t$, thus a different symbol, $\tau$, is used.  

Various approaches have been proposed to improve the computational efficiency of the original LSS method.  The Multiple Shooting Shadowing (MSS) algorithm \cite{BLONIGAN2018447} was introduced to reduce the memory requirements of the standard LSS. When coupled with a matrix-free pre-conditioner to improve the convergence rate \cite{SHAWKI2019108861}, MSS has lower computational cost and memory requirements than standard LSS. In \cite{NI201756}, the non-intrusive LSS (NILSS) approach was derived and applied successfully to the 2D flow over a backward facing step and later to 3D flows inside a channel and around a cylinder \cite{BLONIGAN2017803,Ni_2019}. The computational cost of NILSS methods scales linearly  with the number of positive Lyapunov exponents (PLEs). 
Theoretical predictions indicate that the highest Lyapunov exponent scales with the inverse of the Kolmogorov time scale \cite{Crisanti_et_al_1993,Hassanaly_Raman_2019}. Yet another approach is the formulation of the shadowing problem in the frequency domain; to this end, the shadowing harmonic operator was introduced recently \cite{KANTARAKIAS2023111757}. 
The cost of the method is case-dependent, but for the Kuramoto-Sivashinsky system, sensitivities are computed at a cost roughly equal to that of the baseline solution.

In this paper, we derive the adjoint of the shadowing harmonic operator. The adjoint formulation allows us to compute the sensitivity of a time-average QoI of a chaotic dynamical system with respect to multiple parameters at a cost independent of the number of parameters. These sensitivities are then used to compute the UQ spectral coefficients in the context of se-gPC. As mentioned earlier, for spectral order $p = 1$, the cost of se-gPC is independent of $m$. This is the first application of UQ in chaotic systems with computational cost independent of the number of stochastic parameters. The method is applied and tested to the stochastically forced Kuramoto-Sivashinski equation and the results are compared against reference data obtained from Monte Carlo simulations. 

The rest of this paper is structured as follows: in section \ref{sec:2} an overview of the standard gPC and se-gPC formulations is given. The adjoint shadowing harmonic operator for a general dynamical system is derived in section \ref{sec:3}.  The se-gPC is applied to the Kuramoto-Sivashinsky equation with a single and multiple uncertain forcing parameters in sections \ref{sec:4} and \ref{sec:5} respectively. Finally, in section \ref{sec:6} the main findings of the paper are summarized.

\section{\label{sec:2}Sensitivity-Enhanced Uncertainty Quantification}
Consider a dynamical system governed by a set of ordinary differential equations, 
\begin{equation}
\label{eq:1_ODE}
\begin{aligned}
&\frac{d \bu}{dt} = f(\bu;\bs)\\
&\bu(0; \bs) = \bu_0(\bs)
\end{aligned}
\end{equation}
\noindent where $\bu(t;\bs) \in \mathbb{R}^{N_{\bu}}$ is the vector of state variables and $\bs \in \mathbb{R}^{N_{\bs}}$  is a set of control parameters that define the dynamics of the system (for example Reynolds number in the case of incompressible fluid flows). We assume that the vector field $ \bf: \mathbb{R}^{N_{\bu}} \times  \mathbb{R}^{N_{\bs}} \to \mathbb{R}^{N_{\bu}}$ varies smoothly with $\bu$ and $\bs$. 
In most practical applications, we are interested in a time-averaged quantity $\overline{J}(\bs): \mathbb{R}^{N_{\bs}} \to \mathbb{R}$, 
\begin{equation}
\label{eq:2_QoI}
    \overline{J}(\bs) = \lim_{T \to \infty} \frac{1}{T} \int_0^T J(\bu,\bs) dt,
\end{equation}
which is usually referred to as the Quantity of Interest (QoI), for example lift or drag coefficient of an aerofoil. We assume that the control parameters $\bs$ are functions of $m$ independent stochastic variables  $\ksi_i$ that form the vector  $\vksi = [ \ksi_1, \dots, \ksi_m]$. Each random variable $\ksi_i$ is characterized by a probability density function (PDF), $w_i(\ksi_i)$ in the domain $\mathcal{E}_i$. We seek to estimate the effect of  $\vksi$ to the statistics of $\overline{J}(\bs)$. 

This effect can be quantified via the generalized polynomial chaos (gPC) expansion. In gPC a complete probability space $\mathbb{P} = (\Omega, \Sigma,d\mathcal{P})$ is defined, where  $\Omega$ refers to the set of random events and the probability measure $d\mathcal{P}$ is characterized by the $\sigma$-algebra $\Sigma$.
The vector $\vksi$ follows the PDF $W = \prod_{i=1}^m w_i({\ksi_i})$, defined in the domain $\mathcal{E} = \prod_{i=1}^m \mathcal{E}_i $.
This stochastic space is spanned by a polynomial basis $\Psi \!=\! \{ \Psi_0, \Psi_1, \dots \}$ which is orthogonal to $W$  with respect to the inner product, 
\begin{equation}
\langle \Psi_j,\Psi_k \rangle = \int_{\mathcal{E}} \Psi_j \Psi_k W d\vksi = \delta_{jk } \langle \Psi_j,\Psi_j \rangle. 
\label{theory01}
\end{equation}
The polynomial basis is normalized so that $\left <\Psi_j,\Psi_j \right > = 1$. When $m>1$, $\Psi$ is defined by the tensor product of the unitary polynomials $\psi^{(i)}$, as in $\herm := \otimes_{i=1}^m \psi^{(i)} = \{\herm_0,\herm_1,\dots \}$.
The base is truncated to a finite number of polynomials by limiting the order of $\Psi_j$ to $p$.
In that case, the QoI $\overline{J}$ is written in spectral form as
\begin{equation}
\overline{J} (\vksi) =\sum_{i=0}^{P} c^{(i)} \herm_i(\vksi) + \boldsymbol{\epsilon}(\vksi), 
\label{eq:pce01}
\end{equation}
where the number of basis functions is given by,
\begin{equation}
P+1 = \frac{(p+m)!}{p! m!},
\label{theory03}
\end{equation}
and $\boldsymbol{\epsilon}(\vksi)$ is the truncation error (due to finite $P$). In UQ with gPC, the goal is to compute the spectral coefficients $c^{(i)}$ in a computationally efficient manner. The moments of $\overline{J} (\vksi)$ can be easily computed algebraically from $c^{(i)}$. 

In this paper, the coefficients are computed via a Weighted Least Squares (WLS) approach. To this end, $q$ realizations of $\vksi$ are defined, with the $i$-th realization written as $\vksi^{(i)}= [ \ksi_1^{(i)}, \dots, \ksi_m^{(i)}]$. The QoI $\overline{J}$ is computed for $q$ realizations and stored in the vector $\mathbf{Q} = [\overline{J}(\vksi^{(1)}), \dots,  \overline{J}(\vksi^{(q)})]^{\top}$.
Defining the vector $\boldsymbol{c} = [c^{(0)}, \dots, c^{(P)} ]^\top \in \mathbb{R}^{P+1}$, equation \eqref{eq:pce01} can be written in  matrix form as,
\begin{equation}
\mathbf{Q}  = \boldsymbol{\psi} \boldsymbol{c} + \boldsymbol{\epsilon}, 
\label{eq:pce04}
\end{equation}
where $\boldsymbol{\psi}$ is the measurement matrix with elements $\boldsymbol{\psi}_{ij} = \herm_j(\vksi^{(i)})$, 
i.e.\ the $i$-th row contains the values of the orthogonal polynomial basis $\herm_j$ evaluated at the $i$-th sample point $\vksi^{(i)}$, and $\boldsymbol{\epsilon}$ is the vector of truncation errors. The spectral coefficients are computed by solving the following weighted least squares minimization problem,
\begin{equation}
\min_{\boldsymbol{c}}\| \boldsymbol{W}^{\frac{1}{2}} \left (  \mathbf{Q} - \boldsymbol{\psi} \boldsymbol{c}  \right )  \|_2^2=
\min_{c}  \left(\mathbf{Q}- \boldsymbol{\psi} \boldsymbol{c}\right)^\top  \boldsymbol{W} \left(\mathbf{Q}- \boldsymbol{\psi} \boldsymbol{c}\right),
\label{eq:pce05}
\end{equation}
where $\boldsymbol{W} = (\boldsymbol{W}^{\frac{1}{2}})^{\top} \boldsymbol{W}^{\frac{1}{2}}$. The weighting matrix $\boldsymbol{W}^{\frac{1}{2}}$ is a diagonal positive-definite matrix, to be defined later. The solution of \eqref{eq:pce05} results in the normal set of equations, 
\begin{equation}
\left ( \boldsymbol{\psi}^{\top} \boldsymbol{W} \boldsymbol{\psi} \right ) \hat{\boldsymbol{c}} =   \boldsymbol{\psi}^{\top} \boldsymbol{W} \mathbf{Q}.
\label{eq:pce02}
\end{equation}
For system \eqref{eq:pce02} to be well conditioned $q \gg P+1$. Evaluating the QoI $\overline{J}(\vksi^{(i)})$ at the $q$ sample points is computationally expensive, and dominates the cost of the method.  As mentioned in the introduction, when the number of uncertain parameters $m$ is large, the number of spectral coefficients $P$ grows exponentially (equation \eqref{theory03} indicates $P+1 \sim m^p$), leading to large computational cost, known as the 'curse of dimensionality'.

The problem can be mitigated by enriching system \eqref{eq:pce04} with gradient information. This method, called the sensitivity enhanced gPC, or se-gPC, is presented in \cite{Kantarakias_Papadakis_2023}. Differentiating eq. \eqref{eq:pce01} with respect to the $k$-th random variable at the $j$-th sample point we get,
\begin{equation}
\frac {\partial \overline{J}}{\partial \ksi_k^{(j)}}  =\sum_{i=0}^{P} c^{(i)} \frac{\partial \herm_i}{ \partial \ksi_k^{(j)}} +\boldsymbol{\eta}_k(\ksi^{(j)}) \quad (j=1,\dots,q).
\label{eq:sepce0}
\end{equation}
For each random variable $k$, the block of $q$ equations \eqref{eq:sepce0} can be written in matrix form as, 
\begin{equation}
\frac{\partial {\mathbf{Q}}}{\partial \ksi_k}=
\frac{\partial \boldsymbol{\psi} }{\partial \ksi_k}
\boldsymbol{c}+\boldsymbol{\eta}_k \quad (k=1,\dots,m),
\label{eq:sepce02}
\end{equation}
where matrix $\frac{\partial \boldsymbol{\psi} }{\partial \ksi_k}$ contains the gradients of the basis functions,  
\begin{equation}
\frac{\partial \boldsymbol{\psi} }{\partial \ksi_k}=
\begin{bmatrix}
\frac{\partial \Psi_0 \left(\vksi^{(1)}\right) }{\partial \ksi_k^{(1)}}   &  \dots  & \frac{\partial \Psi_P \left(\vksi^{(1)}\right)}{\partial \ksi_k^{(1)}}   \\
  \vdots     &  \ddots & \vdots    \\
\frac{\partial \Psi_0 \left(\vksi^{(q)} \right)}{\partial \ksi_k^{(q)}}   &  \dots   & \frac{\partial \Psi_P \left(\vksi^{(q)} \right)}{\partial \ksi_k^{(q)}}
\end{bmatrix}.\\
\label{eq:nabla_psi_matrix}
\end{equation}
and vector $\frac{\partial {\mathbf{Q}}}{\partial \ksi_k}$ stores the gradient of $\overline{J}$ with respect to the $k$-th random parameter at the $q$ sample points,
\begin{equation}
\frac{\partial {\mathbf{Q}}}{\partial \ksi_k} =\left [\frac{\partial \overline{J}}{\partial \ksi_k^{(1)} }, \dots, \frac{\partial \overline{J}}{\partial \ksi_k^{(q)} } \right]^\top
\label{eq:sepce02_1}
\end{equation}
By stacking together $\mathbf{Q}$ and $\frac{\partial {\mathbf{Q}}}{\partial \ksi_k}$, we define the following block column vector  $\mathbf{G} \in \mathbb{R}^{(1+m)q \times 1}$
\begin{equation}
\mathbf{G}  =  \left [\mathbf{Q}, \frac{\partial {\mathbf{Q}}}{\partial \ksi_1}, \dots, \frac{\partial {\mathbf{Q}}}{\partial \ksi_m} \right ]^{\top}
\label{eq:expanded_G}
\end{equation}
Similarly, by stacking together the measurement matrix $\boldsymbol{\psi}$ and its sensitivity $\frac{\partial \boldsymbol{\psi} }{\partial \ksi}$, we define the following block matrix $\boldsymbol{\phi} \in \mathbb{R}^{(1+m)q \times (P+1)}$
\begin{equation}
\boldsymbol{\phi}= \left [ \boldsymbol{\psi}, \frac{\partial \boldsymbol{\psi} }{\partial \ksi_1}, \dots, \frac{\partial \boldsymbol{\psi} }{\partial \ksi_m}  \right ]^{\top}
\label{eq:expanded_phi}
\end{equation}
We can therefore write the following $(1+m)\times q$ equations for the spectral coefficients $\boldsymbol{c}$,
\begin{equation}
\mathbf{G}  = \boldsymbol{\phi} \boldsymbol{c} +\boldsymbol{\theta}.
\label{eq:se_pce04}
\end{equation}
As before, to compute the coefficients $\boldsymbol{c}$, we solve the following minimization problem
\begin{equation}
\min_{c}  \| \boldsymbol{W}'^{\frac{1}{2}} \left(\mathbf{G}- \boldsymbol{\phi} \boldsymbol{c}\right) \|_2^2 =\min_{c}  \left(\mathbf{G}- \boldsymbol{\phi} \boldsymbol{c}\right)^\top  \boldsymbol{W}' \left(\mathbf{G}- \boldsymbol{\phi} \boldsymbol{c}\right),
\label{eq:minimization_problem_segPC}
\end{equation}
where $\boldsymbol{W}'$ is a block diagonal weighting matrix, consisting of $1+m$ blocks $\boldsymbol{W}$. The solution $\hat{\boldsymbol{c}}$ is obtained via the normal equations, 
\begin{equation}
\left( \boldsymbol{\phi}^{\top}   \boldsymbol{W}' \boldsymbol{\phi} \right) \hat{\boldsymbol{c}} =   \boldsymbol{\phi} ^{\top}  \boldsymbol{W}' \mathbf{G},
\label{eq:sepce5}
\end{equation}
Notice the similarity between equations \eqref{eq:pce02} and \eqref{eq:sepce5}. The weights $\boldsymbol{W}^{\frac{1}{2}}$ are computed with asymptotic sampling, a version of coherence  sampling, see \cite{HAMPTON201573}. In this paper for simplicity (and without loss of generality)  only Gaussian inputs are considered, and the weights can be computed analytically as,
\begin{equation}
W^{\frac{1}{2}}_{ii}(\vksi) = exp(-\| \vksi \|^2/4)
\label{eq:sepce6}
\end{equation}
For more details on the weights calculation and for extension to other input distributions, see \cite{HAMPTON201573}.  


To avoid large values of $q$, and thus keep the computational cost low, it is important to sample the QoI effectively. Different algorithms to sample the stochastic space are presented in \cite{HADIGOL2018382}. In this paper, we apply QR decomposition. This is a greedy algorithm that maximises the determinant of a matrix; in this sense it is a D-optimal design method, see \cite{HADIGOL2018382,SOMMARIVA20091324,BruntonPaper}. The process is as follows: A large pool of $q$ random sample points is  generated (from the prescribed probability density functions) and the measurement matrix $\boldsymbol{\psi}$ is formed. The question is how to select a subset of at least $P+1$ points from this large pool. To this end, we multiply equation \eqref{eq:pce04} with the row selection matrix $\pvt \in \mathbb{R}^{(P+1)\times q}$, thus we have,
\begin{equation}
{\pvt } \mathbf{Q} = \pvt \boldsymbol{\psi} {\boldsymbol{c}}. 
\label{eq:tailored04b}
\end{equation}
At each row of $\pvt$ all elements are $0$, except the element at the column that corresponds to the selected sampling point, which takes the value of $1$. In D-Optimal experiment design, $\pvt$ is found as a solution to the following maximization problem,
\begin{equation}
\pvt =\underset{\boldsymbol{P}}{\mbox{argmax}} \mbox{ }  \left |  det \left(\pvt  \boldsymbol{W}^{1/2} \boldsymbol{\psi}\right)  \right | 
\label{eq:tailored06}
\end{equation}
where $det ()$ denotes the determinant of a matrix. The solution to this problem is given via the pivoted QR decomposition, 
\begin{equation}
 \left( \boldsymbol{W}^{\frac{1}{2}} \boldsymbol{\psi} \right)^\top \pvt = {\boldsymbol{Q}} {\boldsymbol{R}}
\label{eq:tailored01}
\end{equation}
The index matrix $\pvt$ is chosen so that the diagonal elements $r_{ii}$ of ${\boldsymbol{R}}$ are ranked in descending magnitude $|r_{11}| \geq |r_{22}| \geq \dots \geq |r_{ii}|$. 

Since each sample point offers $1+m$ equations, at least  $\frac{P+1}{1+m}$ samples with the highest $r_{ii}$ scores are retained. Thus sensitivity enhancement reduces the computational cost compared to standard gPC by a factor $m$ at the cost an adjoint evaluation at each sample point. 

We could have applied the same approach to system  \eqref{eq:se_pce04} directly. However, applying QR decomposition to matrix  $\left( \boldsymbol{W}^{\frac{1}{2}} \boldsymbol{\psi} \right)^\top $ and computing at these points the QoI and its gradient is preferable compared to QR decomposition of the weighted augmented matrix $\left( \boldsymbol{W}'^{\frac{1}{2}} \boldsymbol{\phi} \right)^\top$,  see \cite{Kantarakias_Papadakis_2023} for a comparison between the two approaches.

\section{\label{sec:3} Adjoint Sensitivity Analysis of Chaotic Systems}
In this section we present a method for the computation of sensitivities of time-average quantities of chaotic systems to multiple parameters. To this end, we derive the adjoint version of the shadowing harmonic operator introduced in \cite{KANTARAKIAS2023111757}.

The goal of the LSS is to find a shadowing trajectory at $\bs + d\bs$ that stays in close proximity (i.e. shadows) the reference trajectory at $\bs$. If the underlying system is uniformly hyperbolic, the shadowing trajectory is guaranteed to exist. To achieve this goal, LSS  solves the following minimization problem, see \cite{BLONIGAN2018447},
\begin{subequations}
\begin{align}
& \min_{\bv,\eta}\frac{1}{2} \int_0^T  \| \bv(t) \|^2 dt  \mbox{  } s.t.  \label{eq:cost_function_3} \\
&\frac{d \bv}{dt} = \frac{\partial \bf}{\partial \bu} \bv + \frac{\partial \bf}{\partial \bs} + \eta(t) \bf \label{eq:4_ODE_tangent_dilation}\\
& \langle  \bf(\bu;\bs) \bv(\bu;\bs) \rangle  = 0,
\label{eq:orthogonality}
\end{align}
\label{eq:minimisation_3}
\end{subequations}
\noindent where $\bv(t) = \frac{d\bu (\tau(t); \bs)}{ d \bs}$ is the sensitivity of the solution $\bu(t; \bs)$ to a change $\delta \bs$ of $\bs$, $\eta(t) = \frac{d}{d \bs} \left( \frac{d \tau}{dt} \right)$ is the time-dilation, while \eqref{eq:orthogonality} denotes the orthogonality between the vectors $\bf(\bu;\bs)$ and $\bv(\bu;\bs)$ at each point along the trajectory.  The gradient $\frac{d \overline{J}}{d \bs}$ is then given by the following expression,
\begin{equation}
\label{eq:5_sensitivity_expression}
\frac{d \overline{J}}{d \bs} = \frac{1}{T} \int_0^T \frac{\partial J}{\partial \bu} \bv + \frac{\partial J}{\partial \bs} + \eta(t) (J - \overline{J}) dt
\end{equation}
The solution of \eqref{eq:minimisation_3} has shown to produce accurate sensitivities \cite{BLONIGAN201416,BLONIGAN2017803,WangLSS,WANG2014210}.
 
In \cite{KANTARAKIAS2023111757} the shadowing operator was formulated  in the frequency domain. The key idea is to replace the minimization \eqref{eq:cost_function_3} with the periodicity condition, yielding the following set of equations,
\begin{subequations}
\begin{align}
&\frac{d \bv}{dt} = \frac{\partial \bff}{\partial \bu} \bv + \frac{\partial \bff}{\partial \bs} + \eta(t) \bff \label{eq:4_ODE_tangent_dilation 1 }\\
& \langle  \bff(\bu;\bs) \bv(\bu;\bs) \rangle  = 0  \label{eq:orthogonality chapter 6 1} \\
& \bv(0) = \bv(T) . 
\label{eq:cost_function_3 chapter 6 1} 
\end{align}
\label{eq:minimisation_3 chapter 6 1}
\end{subequations}
The new contribution of the present paper with respect to \cite{KANTARAKIAS2023111757} is that an adjoint approach is taken, since sensitivities with respect to a large number of parameters are required. 
To this end, a Lagrangian function is defined, 
\begin{equation}
\label{eq:5_Lagrangian1}
\mathcal{L } =  \frac{d \overline{J}}{d \bs} + \frac{1}{T} \int_0^T \lambda^\top \mathcal{R}_{\bv} dt + \frac{1}{T} \int_0^T \mu \mathcal{R}_{\eta} dt,
\end{equation}
where $\mathcal{R}_{\bv} \in \mathbb{R}^{N_{\bu}}$ and $\mathcal{R}_{\eta}\in \mathbb{R}$ are the residuals of \eqref{eq:4_ODE_tangent_dilation 1 } and \eqref{eq:orthogonality chapter 6 1}, while $\lambda \in \mathbb{R}^{N_{\bu}}$ and $\mu \in \mathbb{R}$ are the adjoint state variables. This is expanded as, 
\begin{equation*}
\label{eq:5_Lagrangian2}
\begin{aligned}
\mathcal{L } =  &\frac{1}{T} \int_0^T \frac{\partial J}{\partial \bu} \bv + \frac{\partial J}{\partial \bs} + \eta(t) (J - \overline{J}) dt+ \\
&\frac{1}{T} \int_0^T \lambda^\top \Big ( \frac{d \bv}{dt} - \frac{\partial \bff}{\partial \bu} \bv + \frac{\partial \bff}{\partial \bs} + \eta \bff  \Big ) dt + \\
&\frac{1}{T} \int_0^T \mu \left ( \bff^\top \bv \right) dt, 
\end{aligned}
\end{equation*}
and using integration by parts, 
\begin{equation*}
\label{eq:5_Lagrangian3}
\begin{aligned}
\mathcal{L } =  &\frac{1}{T} \int_0^T  \Big ( -\frac{d \lambda^\top}{dt} - \frac{\partial \bff}{\partial \bu} \lambda^\top + \frac{\partial J}{\partial \bu}  + \mu \bff^\top \Big ) \bv dt +\\
&\frac{1}{T} \int_0^T \eta \Big ( J - \overline{J} - \bff^\top \lambda \Big ) dt +  [\bv^\top \lambda]_0^T +\\
&\frac{1}{T} \int_0^T \Big ( \frac{\partial J}{\partial \bs} - \lambda^\top \frac{\partial \bff}{\partial \bs}  \Big ) dt
\end{aligned}
\end{equation*}
We seek to make the Lagrangian independent of $\bv(t)$ and $\eta(t)$. This is achieved by solving the following field adjoint equations,
\begin{subequations}
\begin{align}
&\frac{d \lambda}{dt} = - \left(\frac{\partial \bf}{\partial \bu} \right)^\top \lambda + \frac{\partial J}{\partial \bu} + \mu \bf \label{eq:adjoint_equation} \\
& \bf^\top \lambda  = J - \overline{J} \label{eq:adjoint_orthogonality} \\
& \lambda(0)=\lambda(T) \label{eq:adjoint_periodicity}
\end{align}
\label{eq:6_FAE}
\end{subequations}
The gradients can be computed from,
\begin{equation}
\label{eq:7_SDs}
\frac{d \overline{J}}{d \bs} = \frac{1}{T} \int_0^T \Big ( \frac{\partial J}{\partial \bs} - \lambda^\top \frac{\partial \bf}{\partial \bs}  \Big ) dt.
\end{equation}
Notice that \eqref{eq:adjoint_equation} is similar to  \eqref{eq:4_ODE_tangent_dilation}; the difference is that the transpose of the Jacobean is used and the time dilation term $\eta \bf$ is replaced by the adjoint term $\mu \bf$. The adjoint normality constraint \eqref{eq:adjoint_orthogonality} has as forcing the residual $J - \overline{J}$. Note also that the periodicity condition \eqref{eq:cost_function_3 chapter 6 1} for $\bv$ extends also to $\lambda$, see \eqref{eq:adjoint_periodicity}. 

The above equations can be formulated in the frequency domain by expanding $\lambda(t)$ and $\mu(t)$ in Fourier series as, \begin{equation}
\label{eq:Fourier_Expansions}
\lambda(t) = \sum_{ k = -\ell}^{\ell} \hat{\lambda}_{k} e^{i k\omega_0  t}, \mbox{   } \mu(t) = \sum_{ k = -\ell}^{\ell} \hat{\mu}_{k}  e^{i k\omega_0 t},
\end{equation}
where $\hat{\lambda}_{k},\hat{\mu}_{k}$ are the Fourier coefficients and  $\omega_0 = \frac{2 \pi}{T}$ is the fundamental frequency. The index $k \in [-\ell,\ell]$ denotes the harmonics with frequencies $\omega_k=k\omega_0$.
Introducing expansions \eqref{eq:Fourier_Expansions} into  \eqref{eq:adjoint_equation} and \eqref{eq:adjoint_orthogonality}, the system that yields the Fourier coefficients is written in compact form as 
\begin{equation}
i k\omega_0 \mathcal{I}_u
\begin{bmatrix}
  \hat{\lambda}_{k} \\
 \hat{\mu}_{k}
\end{bmatrix}  
+ \sum_{l= -\ell}^\ell T_{k-l}  
\begin{bmatrix}
  \hat{\lambda}_{l} \\
 \hat{\mu}_{l}
\end{bmatrix}  
= 
\begin{bmatrix}
  \widehat{\frac{d J}{d \bu}}_k \\
 \hat{J_k} - \overline{J} 
\end{bmatrix}  
\label{eq:FAE_FourierSpace}
\end{equation} 
where
\begin{gather}
 \mathcal{I}_{u}
 =
  \begin{bmatrix}
   \mathcal{I}_{N_u} &   0 \\
   0 &   0 
   \end{bmatrix},
 T_{m}=
  \begin{bmatrix}
  \left( -\widehat{\frac{\partial \bf}{\partial \bu}}\right)_{m}^\top  &
  \hat{\bf}_m \\
   \hat{\bf}_{m}^\top  &
   0 
   \end{bmatrix}.
   \label{eq:matrix_T_m}
\end{gather}

We define the block diagonal matrix, 
\begin{equation}
\mathcal{T} (T_{m}) = 
\begin{bmatrix}
T_0   &  T_{-1}  & \dots & T_{-\ell}  &   &  &   \\
T_1  &  T_{0}   & T_{-1} & \dots  & T_{-\ell} &  &  \\
     &  \ddots   & \ddots  & \ddots  &  \dots & \ddots &   \\
T_\ell   &  \dots   & T_{1}  & T_0   & T_{-1} & \dots & T_{-\ell}  \\
     &  \dots    & \ddots  & \ddots   & \ddots  & \ddots & \dots  \\
     &     & T_\ell  & \dots   & T_{1}  & T_{0}  & T_{-1}  \\
          &   &   & T_\ell  & \dots & T_1 & T_0  \\
\end{bmatrix}
,
\label{eq:Block_Toeplitz_matrix}
\end{equation}
which is a block Toeplitz matrix, because each diagonal has the same block. Using $\mathcal{T} (T_{m})$, system \eqref{eq:FAE_FourierSpace} can be written in matrix form as,
\begin{equation}
\left [\mathcal{D} - \mathcal{T} (T_{m}) \right ]  \widehat{\boldsymbol{\Lambda}}
=\widehat{\mathcal{R}},
\label{eq:FAE_system}
\end{equation}
where $\mathcal{D}=diag[ \mathcal{D}_{-q}, \dots, \mathcal{D}_0, \dots, \mathcal{D}_q ]$ is a block diagonal matrix with $\mathcal{D}_k=i k \omega_0  \mathcal{I}_u$, $\widehat{\mathcal{R}} = [ \widehat{\mathcal{R}}_{-q}, \dots, \widehat{\mathcal{R}}_{0}, \dots, \widehat{\mathcal{R}}_{q}]^\top$ with $\widehat{\mathcal{R}}_k = \left [   \widehat{\frac{d J}{d \bu}}_k, \hat{J_k} - \overline{J} \right ]^\top$, and $ \widehat{\boldsymbol{\Lambda}} = \left [ \widehat{\Lambda}_{-q}, \dots, \widehat{\Lambda}_{0}, \dots, \widehat{\Lambda}_{q} \right ]^\top$,  where $\widehat{\Lambda}_k = [\hat{\lambda}_k, \hat{\mu}_k]^\top$. Matrix $\mathcal{H} = \mathcal{D} - \mathcal{T} (T_{m})$ is also known as a Hill matrix. 

Defining the adjoint shadowing harmonic operator as, 
\begin{equation}
\mathcal{A}=
\left [\mathcal{D} - \mathcal{T} (T_{m}) \right ]^{-1},
\label{eq:shadowing_resolvent_operator}
\end{equation}
the solution of system \eqref{eq:FAE_system} can be written symbolically as
\begin{equation}
\widehat{\boldsymbol{\Lambda}}=\mathcal{A} \widehat{\mathcal{R}}.
\label{eq:symbolic_solution}
\end{equation}
The adjoint operator $\mathcal{A}$ maps the forcing $\widehat{\mathcal{R}}$ to the vector $\widehat{\boldsymbol{\Lambda}}$ that contains the unknown Fourier coefficients of the adjoint variables. More details can be found in \cite{WereleyThesis,padovan_otto_rowley_2020,KANTARAKIAS2023111757}. 

We do not directly compute the adjoint shadowing operator $\mathcal{A}$. Instead we apply LU decomposition to the Hill matrix $\mathcal{H} = \mathcal{D} - \mathcal{T} (T_{m})$ and find  $\widehat{\boldsymbol{\Lambda}}$ by forward and back substitution. The gradients can then be computed from, 
\begin{equation}
\frac{d \overline{J}}{ds} = \widehat{\frac{\partial J}{ \partial \bs}}\Bigg |_0 - \sum_{k=-\ell}^{\ell}  \hat{\lambda}_k^\top  \widehat{\frac{\partial \bf}{ \partial \bs}}\Bigg |_{-k},
\label{eq:sensitivity_freq_domain}
\end{equation} 
which is the equivalent of \eqref{eq:7_SDs} in Fourier space.  

The cost of solving system \eqref{eq:FAE_system} is independent of the number of parameters and thus the adjoint formulation can provide the sensitivities of time-average quantities of chaotic systems is a single step. This information is used to augment the se-gPC system as explained in the previous section.

\section{\label{sec:4}Application to the Kuramoto-Sivashinsky system}
The aforementioned methodology is now applied to the forced Kuramoto-Sivashinsky (KS) equation,
\begin{equation}\label{eq:KS_primal}
u_t + u u_x + u_{xx}+ u_{x x x x} = \phi(x) 
\end{equation}
where $x\in[0,L]$ and boundary conditions $u(0,t)=u(L,t)=0 $ and $ u_x (0,t) = u_x(L,t)=0$. Two QoIs are defined, 
\begin{equation} \label{eq:KS_objective}
\begin{aligned}
&\overline{J^{(1)}} = \frac{1}{TL} \int_0^T \int_0^L  J^{(1)}(x,t) dx dt,  \\ 
& \overline{J^{(2)}} = \frac{1}{TL} \int_0^T \int_0^L J^{(2)}(x,t) dx dt, 
\end{aligned}
\end{equation}
where $J^{(1)}(x,t)=u$ and  $J^{(2)}(x,t)=u^2$, that represent time- and space-average values of the state and its energy. The KS equation was discretised in space using second-order accurate central finite differences and integrated with a variable time step Runge-Kutta method. We take $L=128$ that results in chaotic behaviour. The state $u(x,t)$ was stored every $dt = 0.1$ time units and used as input to the adjoint system. The first $1000$ time units were discarded to ensure that the system has reached a chaotic attractor. The next $T = 100$ units were considered as the time horizon for the UQ analysis. 

\begin{figure}
\centering
\includegraphics[scale=0.40, clip]{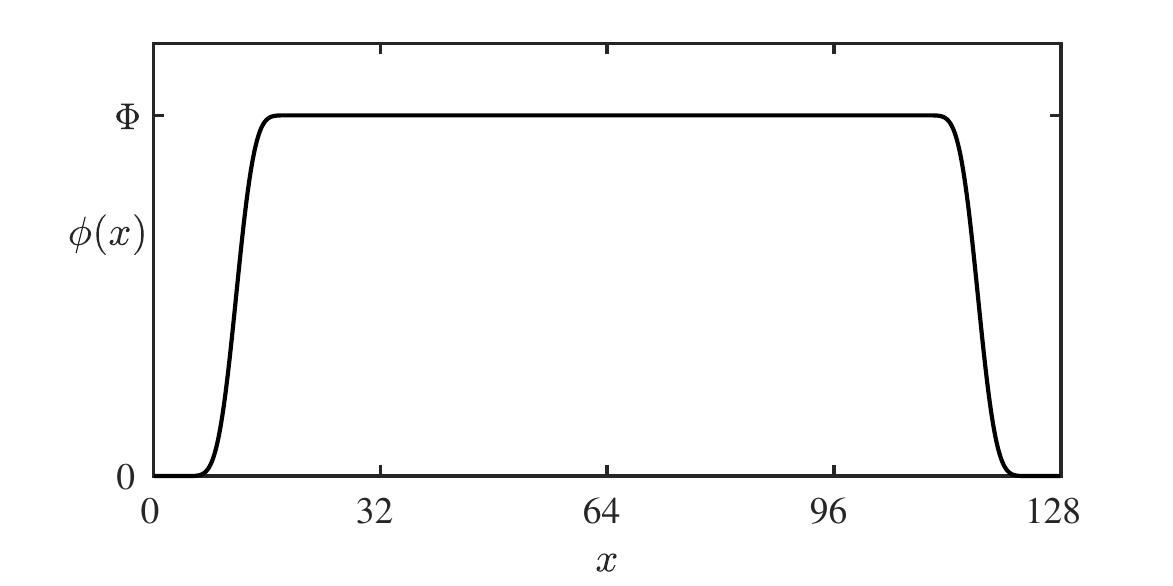}
\caption{Bell-shaped profile $\phi(x)$ with amplitude $\Phi$.}
\label{fig:Bell_Shaped_Input}
\end{figure}

We first consider the bell-shaped forcing distribution $\phi(x)$ shown in figure \ref{fig:Bell_Shaped_Input}, where  $\Phi$ denotes the forcing amplitude. The smooth profiles from $0 \to \Phi$ (close to the left boundary) and from $\Phi \to 0$ (close to the right boundary) are obtained using the error function, $erf$, see \cite{BoydPaper} for more details.  
This forcing is infinitely differentiable and satisfies both Dirichlet and Neumann boundary conditions.  

\begin{figure}
\centering
\begin{subfigure}[b]{0.49\textwidth} 	\includegraphics[scale=0.40, clip]{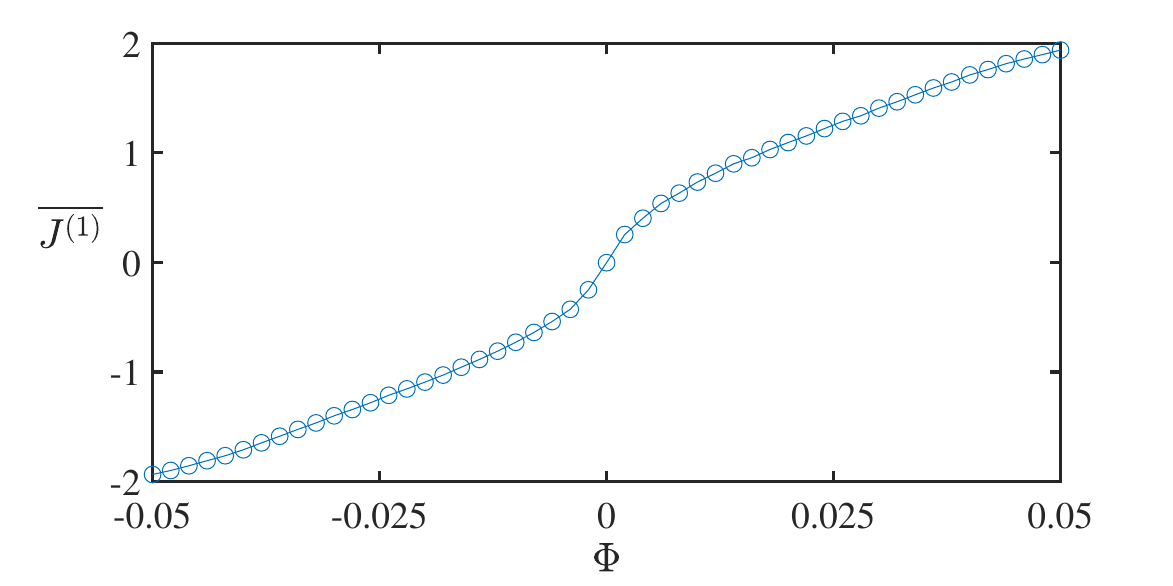}
\caption{}
\label{fig:KS_Functional_Space}
\end{subfigure}
\begin{subfigure}[b]{0.49\textwidth} 	\includegraphics[scale=0.40, clip]{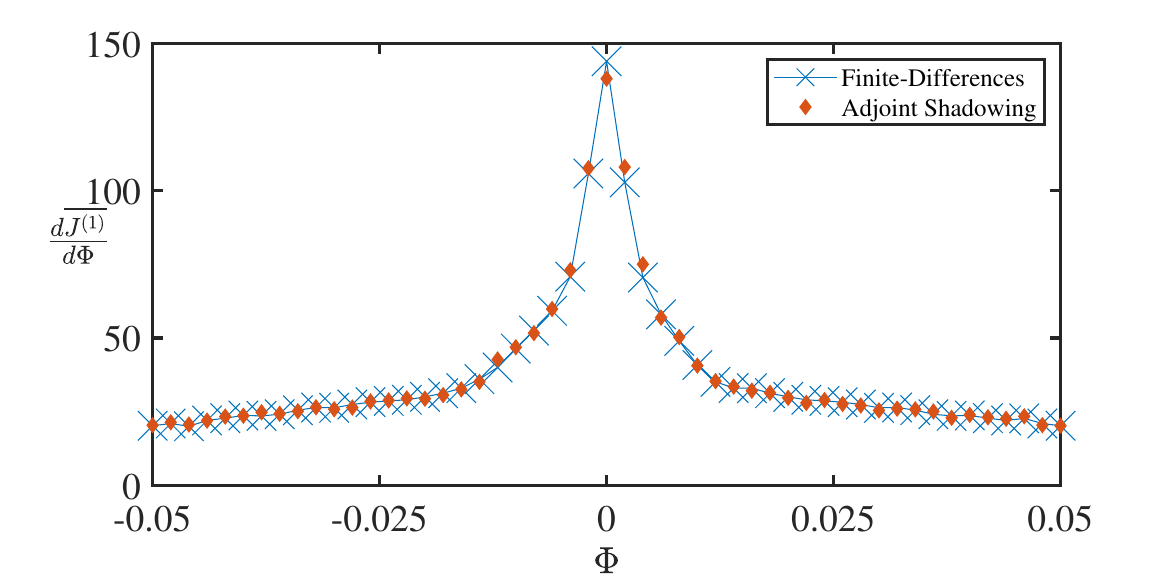}
\caption{}
\label{fig:KS_Functional_Derivative}
\end{subfigure}
\caption{ Variation of (a) $\overline{J^{(1)}}$ with the forcing amplitude $\Phi$ and (b) Sensitivity $\frac{d\overline{J^{(1)}}}{d \Phi}$ computed using the adjoint operator and finite differences. Results are averaged over 200 random initial conditions.}
\label{fig:KS_Uniform_Functional_Space}
\end{figure}

\begin{figure}
\centering
\begin{subfigure}[b]{0.49\textwidth} 	\includegraphics[scale=0.40, clip]{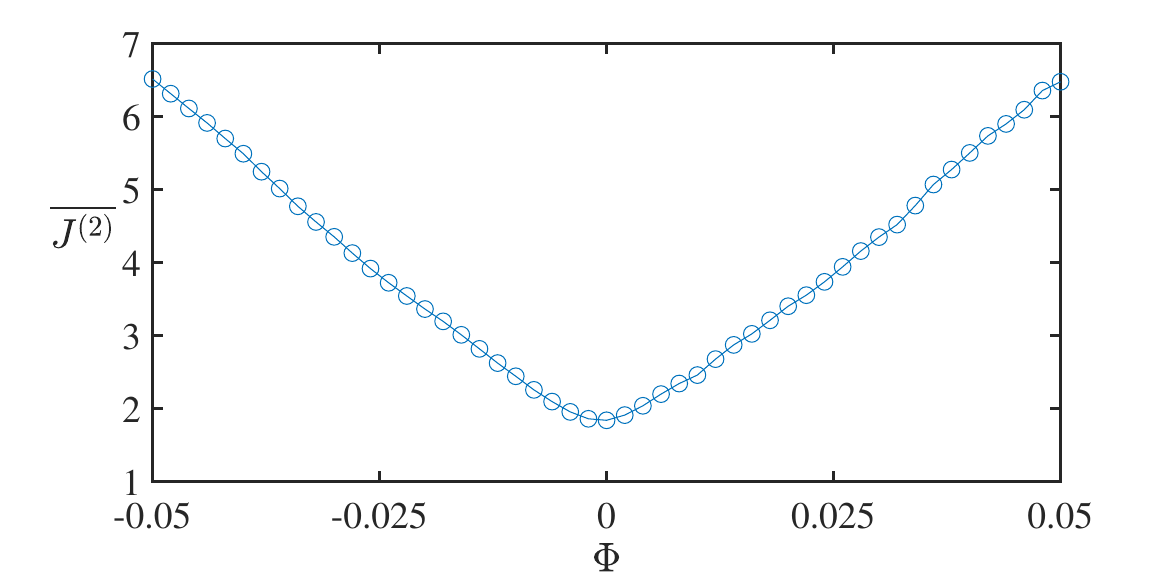}
\caption{}
\label{fig:KS_Functional_Space_u2}
\end{subfigure}
\begin{subfigure}[b]{0.49\textwidth} 	\includegraphics[scale=0.40, clip]{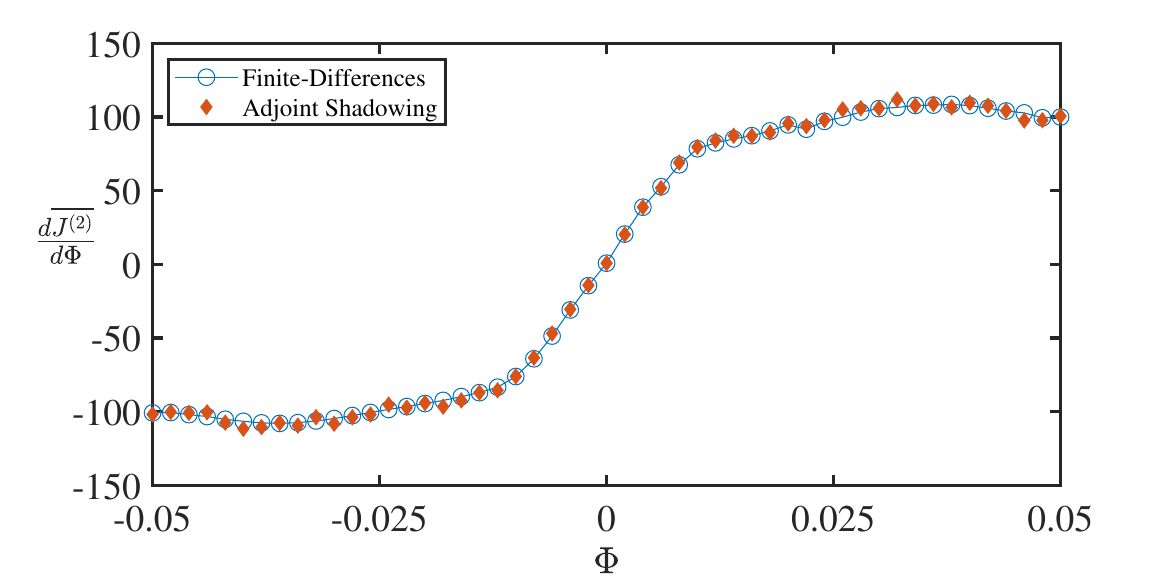}
\caption{}
\label{fig:KS_Functional_Derivative_u2}
\end{subfigure}
\caption{Variation of (a) $\overline{J^{(2)}}$ with the forcing amplitude $\Phi$ and (b) Sensitivity $\frac{d\overline{J^{(2)}}}{d \Phi}$ computed using the adjoint operator and finite differences. Results are averaged over 200 random initial conditions.}
\label{fig:KS_Uniform_Functional_Space_u2}
\end{figure}

The variations of $\overline{J^{(1)}}$ and $\overline{J^{(2)}}$ with respect to amplitude $\Phi$ are shown in figures \ref{fig:KS_Functional_Space} and \ref{fig:KS_Functional_Space_u2} respectively. The sensitivities of these QoIs obtained using finite differences and the adjoint of the shadowing harmonic operator are shown in figures \ref{fig:KS_Functional_Derivative} and \ref{fig:KS_Functional_Derivative_u2} respectively.  
To form the harmonic operator, we consider frequencies in the range $f \in [-0.3,0.3]$ that  captures the active frequency band of the unforced KS system, see \cite{KANTARAKIAS2023111757}. Notice that the two approaches are in very good agreement for both QoIs. In this case, where the sensitivity with respect to a single parameter is considered, the adjoint shadowing operator does not provide any computational advantage over the standard shadowing harmonic  operator, hence this test case is only used as a benchmark to evaluate the accuracy and computational implementation of the method. 

\begin{figure}[ht]
\centering
\begin{subfigure}[b]{0.40\textwidth} 	\includegraphics[scale=0.40, clip]{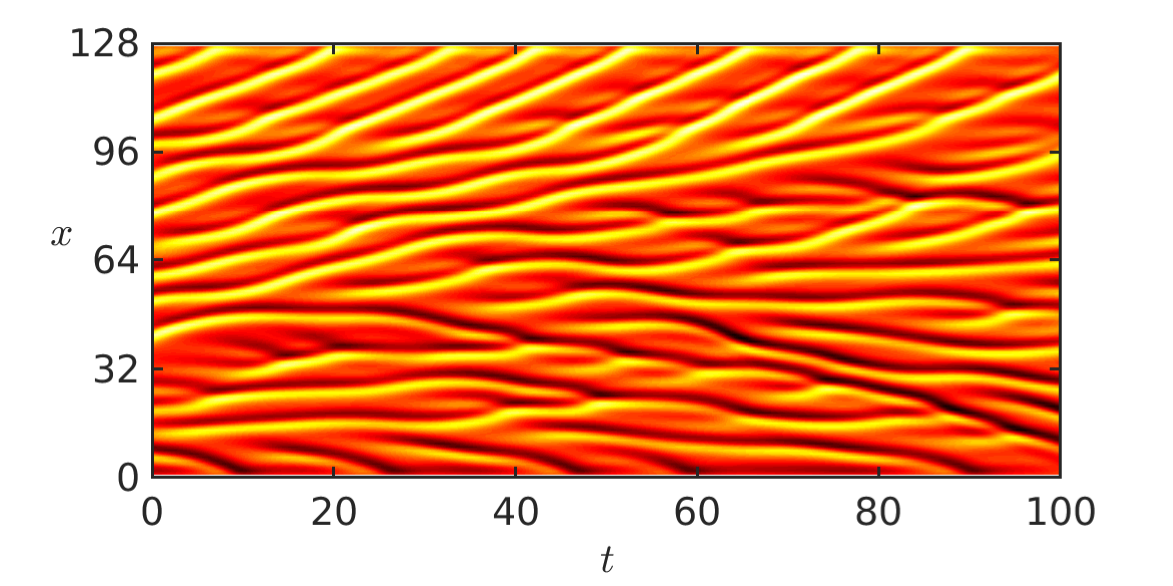}
\caption{}
\label{fig:KS_field_primal}
\end{subfigure}
\begin{subfigure}[b]{0.40\textwidth} 	\includegraphics[scale=0.40, clip]{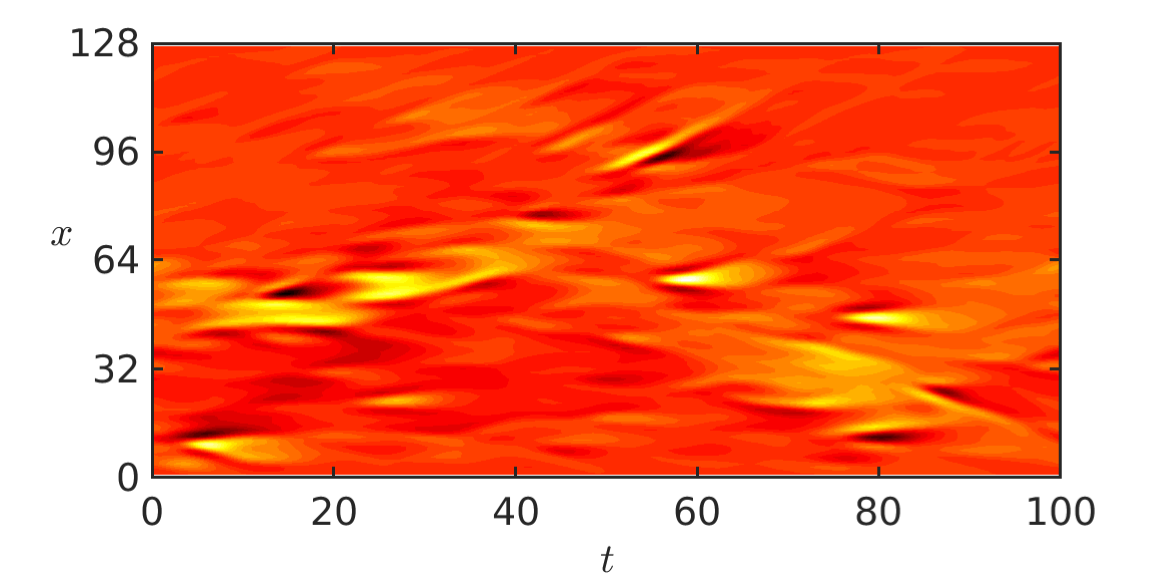}
\caption{}
\label{fig:KS_Field_Adjoint}
\end{subfigure}
\caption{Contour plots of the (a) state $\bu(x,t)$ and (b) adjoint $\lambda(x,t)$ variables for the unforced system, $\phi = 0$. Results from a single realisation with a random initial condition.}
\label{fig:KS_AdjointPrimal_Field}
\end{figure}

Contour plots of the state and adjoint variables in the $(t,x)$ plane for the unforced system, i.e.\ for $\phi(x) = 0$, are shown in figure \ref{fig:KS_AdjointPrimal_Field}. Notice that the adjoint variable $\lambda(x,t)$  does not have the same spatio-temporal streaky structure as the state variable  $\bu(x,t)$. This has been observed before in sensitivity analysis with shadowing method \cite{BLONIGAN2017803}. Similarly to the state $\bu(x,t)$, the spatio-temporal structure of the adjoint state is characterised by high sensitivity to initial conditions. 

We now evaluate the effect of stochastic variation of $\Phi$ to $\overline{J^{(1)}}$ and $\overline{J^{(2)}}$ using the se-gPC method and the sensitivities produced with the adjoint shadowing operator to augment the least squares system. We assume that $\Phi$ follows Gaussian distribution with $\Phi \sim \mathcal{N}(0,\sigma)$, and $\sigma = 0.01$. The standard deviation $\sigma$ is small, but the response of the chaotic system to even small values of $\Phi$ is large, as shown in figures \ref{fig:KS_Functional_Space} and \ref{fig:KS_Functional_Space_u2}. For example, within the range $\Phi \in [-3\sigma, +3\sigma]$, the value of $\overline{J^{(2)}}$ doubles and the sensitivity varies between $-100$ to $+100$; this suggests that the effect of $\Phi$ is strong. In the range of $\Phi$ values considered the system is chaotic. Larger forcing amplitudes were also tested, however they result in forced oscillations with non-chaotic behaviour. 

A Monte-Carlo simulation with $5000$ samples was performed and used as a benchmark to evaluate the accuracy of the se-gPC method. For a single stochastic variable ($m = 1$) and polynomial order $p = 1$, there are $P+1 = 2$ spectral coefficients, while for $p = 2$ there are $P+1 = 3$ coefficients. In the se-gPC we used $q = 4$ samples for $p = 1$, that were augmented with another $4$ equations for the sensitivity of the QoI with respect to $\Phi$. For $p = 2$, $q = 6$ samples were used, augmented with 6 additional equations for the sensitivity. We used more equations than the number of coefficients to account for the (small) variation of the sensitivities to initial conditions. The results for the mean and standard deviation of the QoIs are summarized in table \ref{Comparison_1D_Kuramoto}, where se-gPC is compared with Monte Carlo (MC). There is very good agreement between the two methods. Errors in the standard deviation are less than $1.5\%$ for $p = 1$ and less than $0.5\%$ for $p = 2$ for both QoIs. 

\begin{table}
\begin{tabular}{|l|ll|ll|ll|}
\hline
      & \multicolumn{2}{l|}{se-gPC $p = 1$}         & \multicolumn{2}{l|}{se-gPC $p = 2$}   & \multicolumn{2}{l|}{Monte-Carlo}     \\ \hline \hline
QoI   & \multicolumn{1}{l|}{mean} & std & \multicolumn{1}{l|}{mean} & std & \multicolumn{1}{l|}{mean} & std \\ \hline
$\overline{J^{(1)}}$   & \multicolumn{1}{l|}{$0.0196$}  & $0.6286$     & \multicolumn{1}{l|}{$0.0196$}  & $0.6395$  & \multicolumn{1}{l|}{$0.0195$}  & $0.6370$    \\ \hline
$\overline{J^{(2)}}$    & \multicolumn{1}{l|}{$2.3478$}  & $0.5251$     & \multicolumn{1}{l|}{$2.3505$}  & $0.5174$  & \multicolumn{1}{l|}{$2.3552$}  & $0.5192$     \\ \hline
\end{tabular}
\caption{Comparison between Monte Carlo simulations with $5000$ samples and se-gPC for the KS system. The stochastic input is $\Phi \sim \mathcal{N}(0,0.01)$.}
\label{Comparison_1D_Kuramoto}
\end{table}

\begin{figure}
\centering
\includegraphics[scale=0.40, clip]{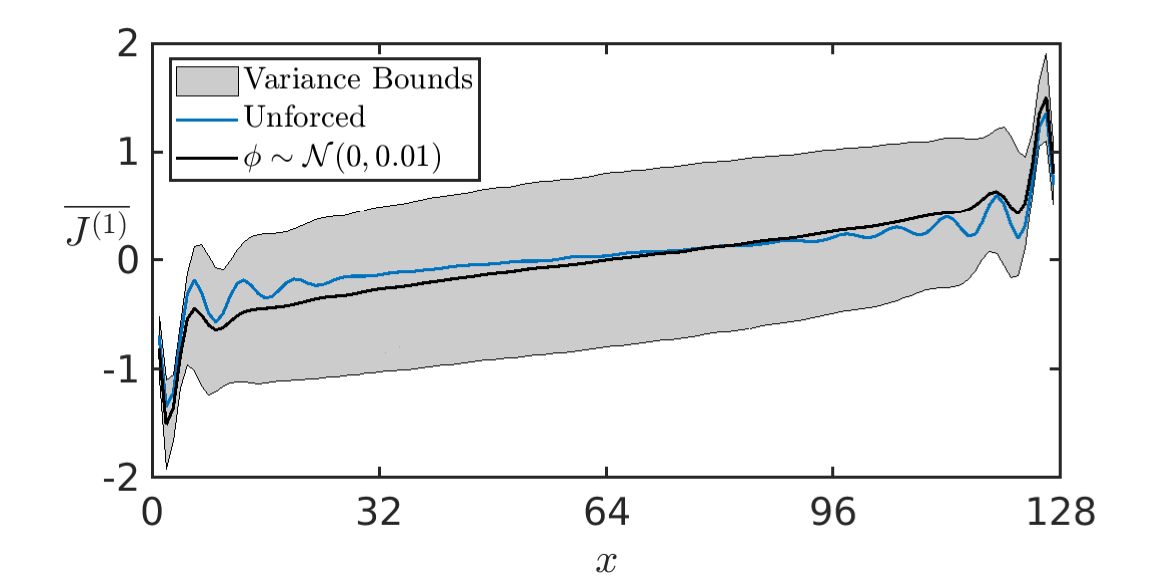}
\caption{$\mathcal{E} \left[\overline{J^{(1)}(x)}\right]$ for the unforced and forced KS system with $\Phi \sim \mathcal{N}(0,0.01)$ for $T = 100$. Results are averaged over 1000 random initial conditions.}
\label{fig:KS_MonteCarlo_Jx}
\end{figure}

To better understand the effect of stochastic forcing $\phi(x)$, in figure \ref{fig:KS_MonteCarlo_Jx} we plot the expectation of the time-average state,  $\mathcal{E}\left[\overline{J^{(1)}(x)}\right]$, where  $\overline{J^{(1)}(x)}=\frac{1}{T} \int_0^T J^{(1)}(x,t) dt=\frac{1}{T} \int_0^T u dt$, against $x$. We compare against the profile of the unforced system ($\phi(x) = 0$), where $\mathcal{E}\left[\overline{J^{(1)}(x)}\right]=\overline{J^{(1)}(x)}$. The results are averaged over $1000$ random initial conditions. 
It's interesting to notice that the stochastic forcing smooths out the spatial oscillations of the time-average state of the unforced system.  We also compute the standard deviation of $\overline{J^{(1)}(x)}$ across $x$, and we superimpose the extent of one standard deviation above and below the expectation (the area between the two boundaries is marked grey). 
The large spread indicates that the actual time-average $\overline{J^1(x)}$ oscillates wildly and can take values much larger that the expectation. Therefore the stochastic forcing drastically affects the output of the system. This explains why the standard deviation of $\overline{J^{(1)}}$ is much larger that the mean (expectation) in table \ref{Comparison_1D_Kuramoto}.  
\section{\label{sec:5} Multidimensional Uncertainty Forcing}
\begin{figure}[ht]
\centering
\includegraphics[scale=0.45, trim={1mm 1mm 1mm 1mm}, clip]{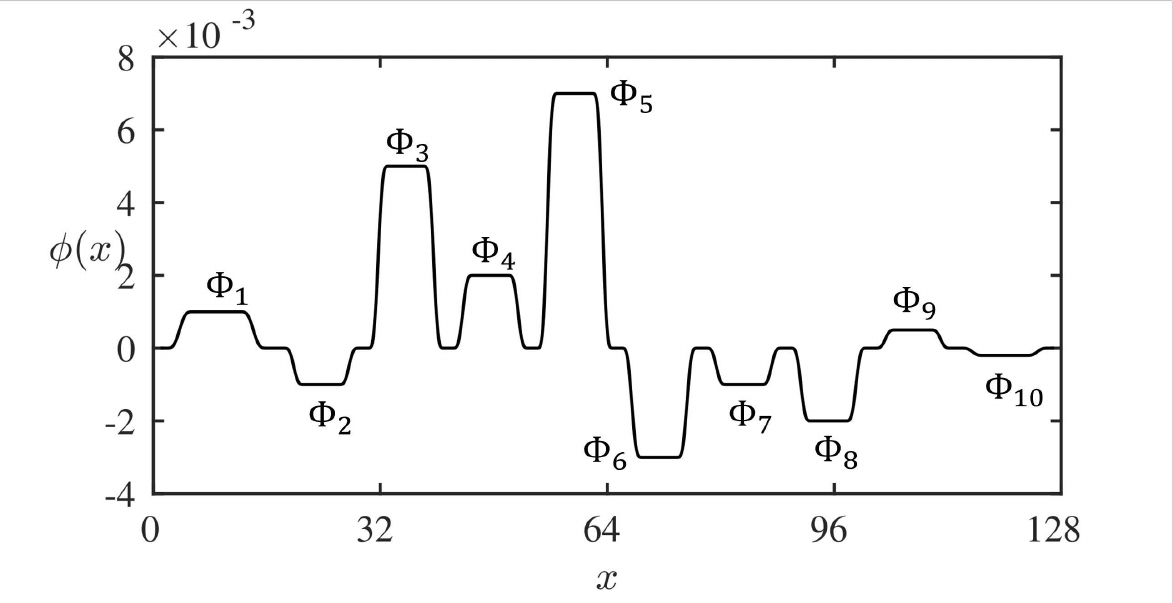}
\caption{Mean forcing $\phi(x)$ with $m = 10$ stochastic parameters, $\Phi_i$ $(i=1, \dots, m$). The forcing amplitudes are $\Phi_1 = 0.001$, $\Phi_2 = -0.001$, $\Phi_3 = 0.005$, $\Phi_4 = 0.002$, $\Phi_5 = 0.007$, $\Phi_6 = -0.003$, $\Phi_7 = -0.001$, $\Phi_8 = -0.002$, $\Phi_9 = 0.0005$, $\Phi_{10} = -0.002$.}
\label{fig:KS_forcing_varied}
\end{figure}

We now consider the stochastic forcing shown in figure \ref{fig:KS_forcing_varied}, which is a  continuous and differentiable profile that contains $10$ peaks and troughs. The local amplitudes $\Phi_i$ are the $m = 10$ independent stochastic variables considered. Such cases are usually found in control problems, where spatially complex forcing allows for more accurate control of the desired output quantities. The mean values of the localised forcing amplitudes are $\Phi_1 = 0.001$, $\Phi_2 = -0.001$, $\Phi_3 = 0.005$, $\Phi_4 = 0.002$, $\Phi_5 = 0.007$, $\Phi_6 = -0.003$, $\Phi_7 = -0.001$, $\Phi_8 = -0.002$, $\Phi_9 = 0.0005$, $\Phi_{10} = -0.002$. These values were selected arbitrarily, since the main objective of the section is to demonstrate the efficiency of the proposed computational approach to conduct UQ. The standard deviations are taken equal to $20\%$ of the mean value, i.e.\ $\sigma_{\Phi_i} = |\Phi_i|/5$. 

\subsection{Characterisation of the adjoint field}
\begin{figure}[ht]
\centering
\begin{subfigure}[b]{0.40\textwidth} 	\includegraphics[scale=0.40, clip]{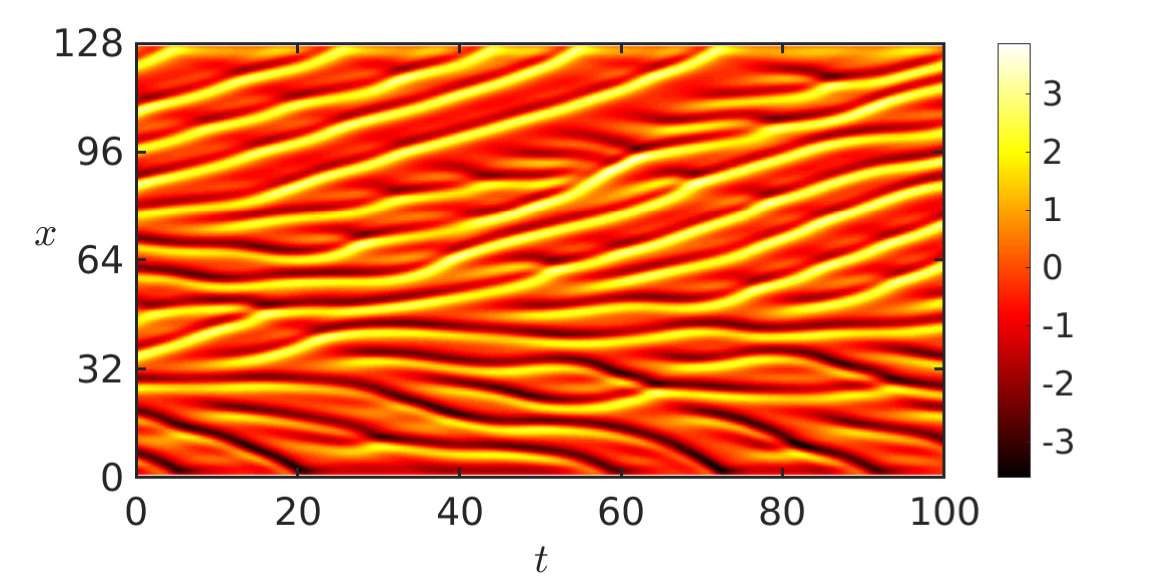}
\caption{}
\label{fig:KS_field_primal_non_uniform_forcing}
\end{subfigure}
\begin{subfigure}[b]{0.40\textwidth} 	\includegraphics[scale=0.40, clip]{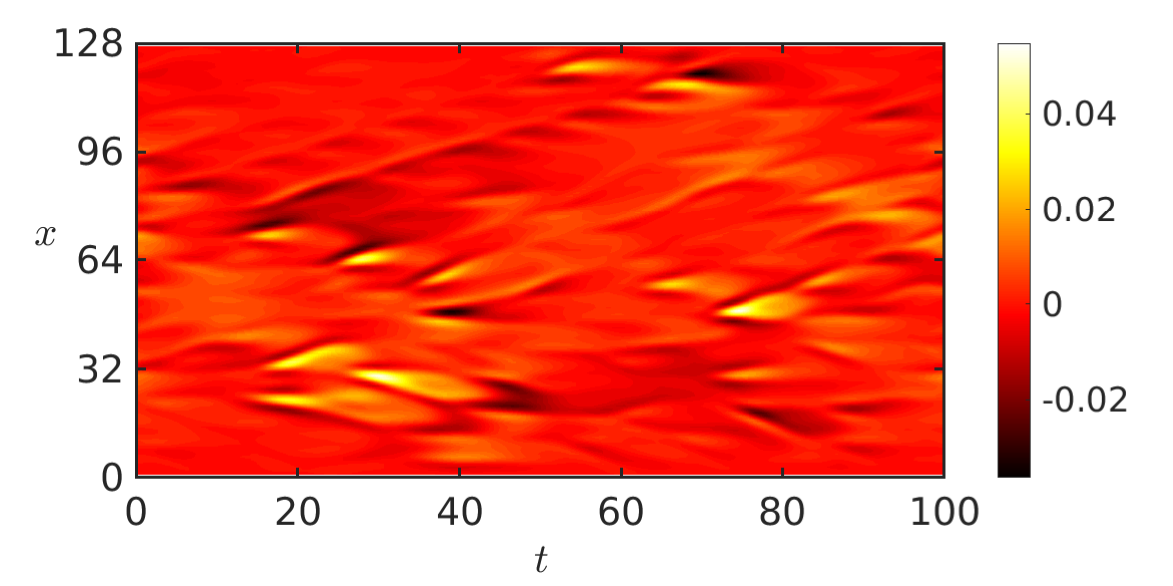}
\caption{}
\label{fig:KS_Field_Adjoint_non_uniform_forcing}
\end{subfigure}
\caption{Contour plots of (a) the state variable and (b) the adjoint variable for the KS system forced with the profile shown in figure \ref{fig:KS_forcing_varied}. Results from a random initial condition.}
\label{fig:KS_AdjointPrimal_Field_non_uniform_forcing}
\end{figure}

Contour plots of the direct and adjoint solutions at the mean values of $\Phi$ are shown in fig. \ref{fig:KS_AdjointPrimal_Field_non_uniform_forcing}. Again the two solutions do not follow a similar structure. The adjoint variable $\lambda(x,t)$ shown in panel (b) maintains small values close to zero, but displays intermittent behaviour with random peaks and troughs that have relatively short time duration.

\begin{figure}[ht]
\centering
\begin{subfigure}[b]{0.40\textwidth} 	\includegraphics[scale=0.40, clip]{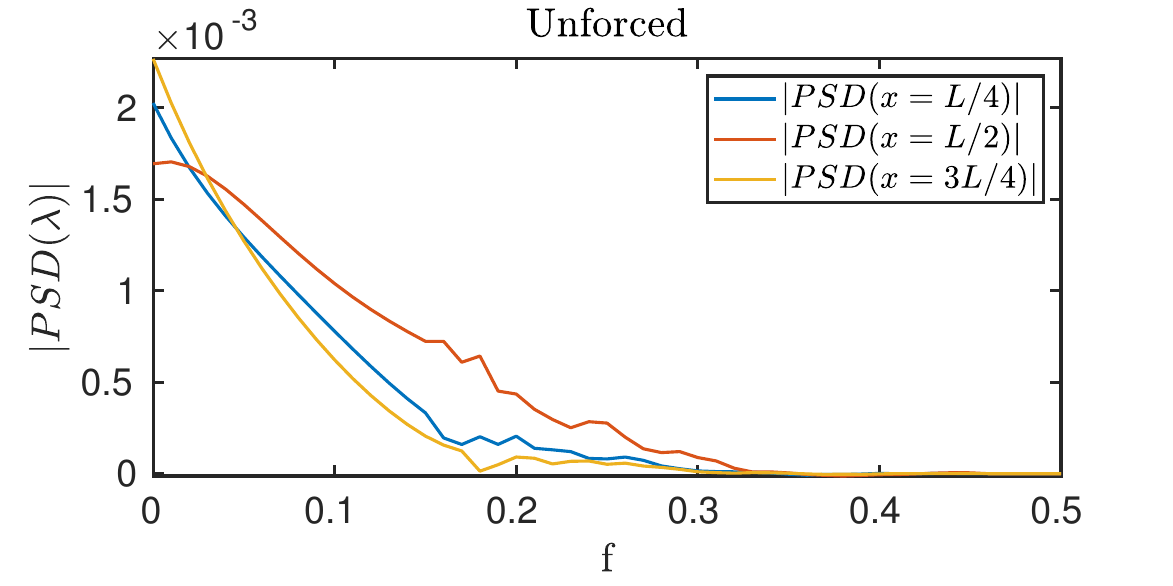}
\caption{}
\label{fig:FFT_Unforced}
\end{subfigure}
\begin{subfigure}[b]{0.40\textwidth} 	\includegraphics[scale=0.40, clip]{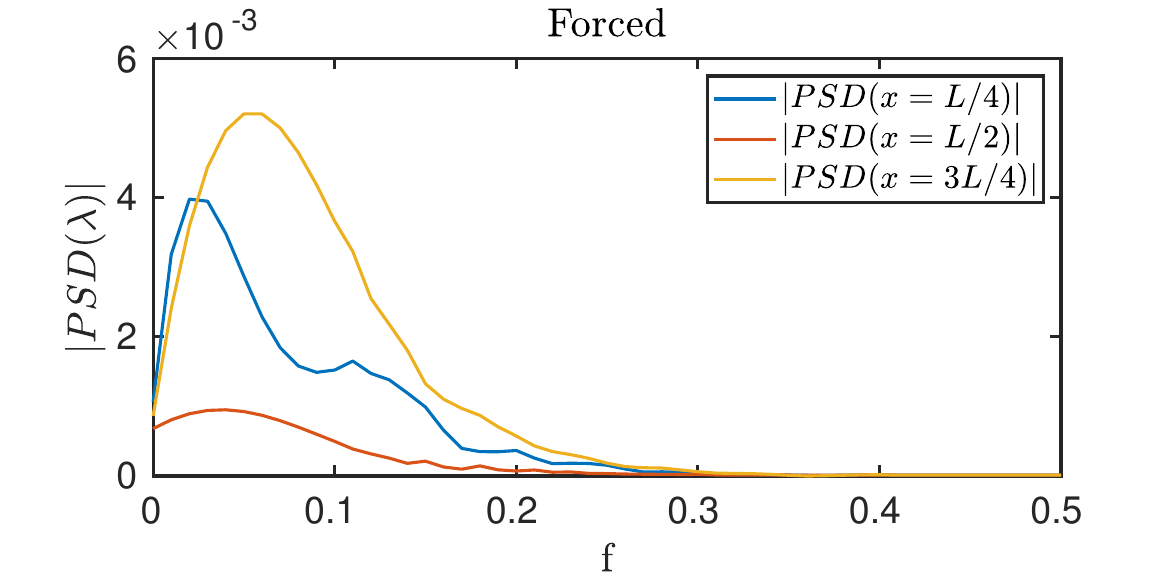}
\caption{}
\label{fig:FFT_Forced}
\end{subfigure}
\caption{Spectra of $\lambda(x,t)$ at  $x = \frac{L}{4}$, $\frac{L}{2}$ and $\frac{3L}{4}$ for (a) the unforced and (b) the forced KS system.}
\label{fig:FFT_Adjoint}
\end{figure}

The spectra of $\lambda(x,t)$ at $x =\frac{L}{4}$, $\frac{L}{2}$ and $\frac{3L}{4}$ are shown in fig. \ref{fig:FFT_Adjoint} for the unforced and forced systems. The spectra were computed with a time window of $T = 100$ and were smoothed with a $5$-th order Savitzky-Golay convolution filter with 5 averaging windows. 
For the unforced system, the spectra are very similar at the three locations.  However, for the forced system the PSD values are larger and vary significantly between the locations due to the spatially varying forcing profile. The lower frequencies are also damped. 

\begin{figure}[ht]
\centering
\includegraphics[scale=0.40, clip]{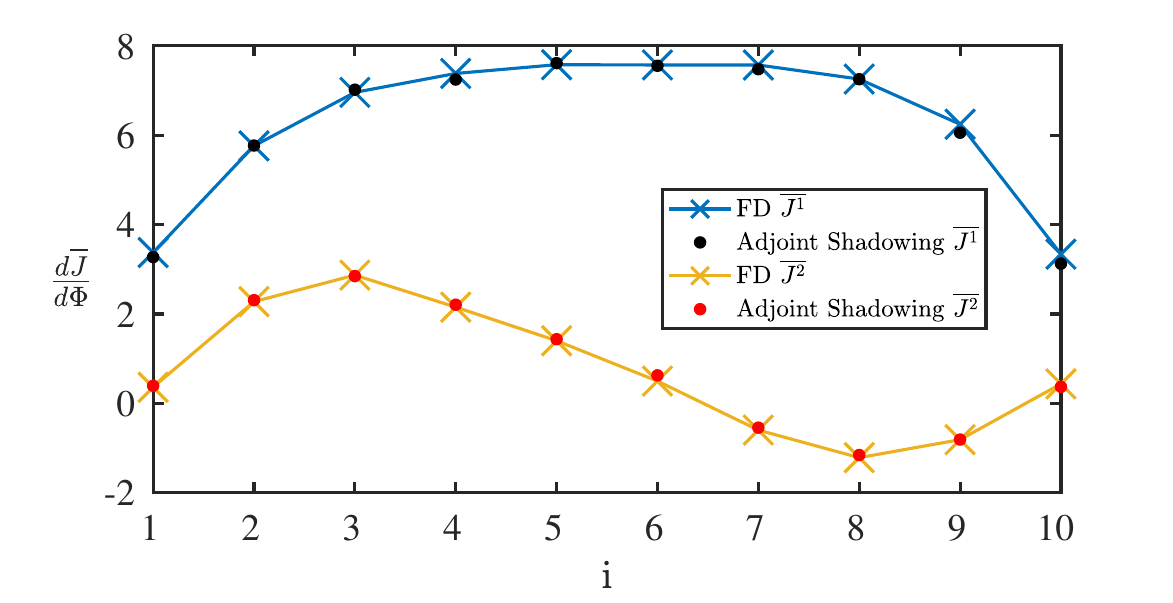}
\caption{Comparison of the sensitivities of $\overline{J^{(1)}}$ and $\overline{J^{(2)}}$ wrt to the amplitudes $\Phi_i$ computed using finite differences and the adjoint shadowing operator. Results are averaged over 100 initial conditions for the adjoint shadowing operator.}
\label{fig:fd_comp_multidimensional}
\end{figure}

The accuracy of the adjoint shadowing harmonic operator is assessed in figure \ref{fig:fd_comp_multidimensional}. The values of the $m = 10$ sensitivities computed by the adjoint operator are compared against reference finite difference results. To evaluate the reference results we varied each $\Phi_i$ separately and averaged over 100 initial conditions (we performed in total 1000 simulations with random initial conditions for all $\Phi_i$'s). The results for the adjoint shadowing operator were averaged over 100 random initial conditions. It is clear from the figure that the adjoint approach computes accurate sensitivities that are in very good agreement with finite differences for both  $\overline{J^{(1)}}$ and $\overline{J^{(2)}}$. 
\subsection{Uncertainty quantification with se-gPC}
\begin{figure}
\centering
\includegraphics[scale=0.40, clip]{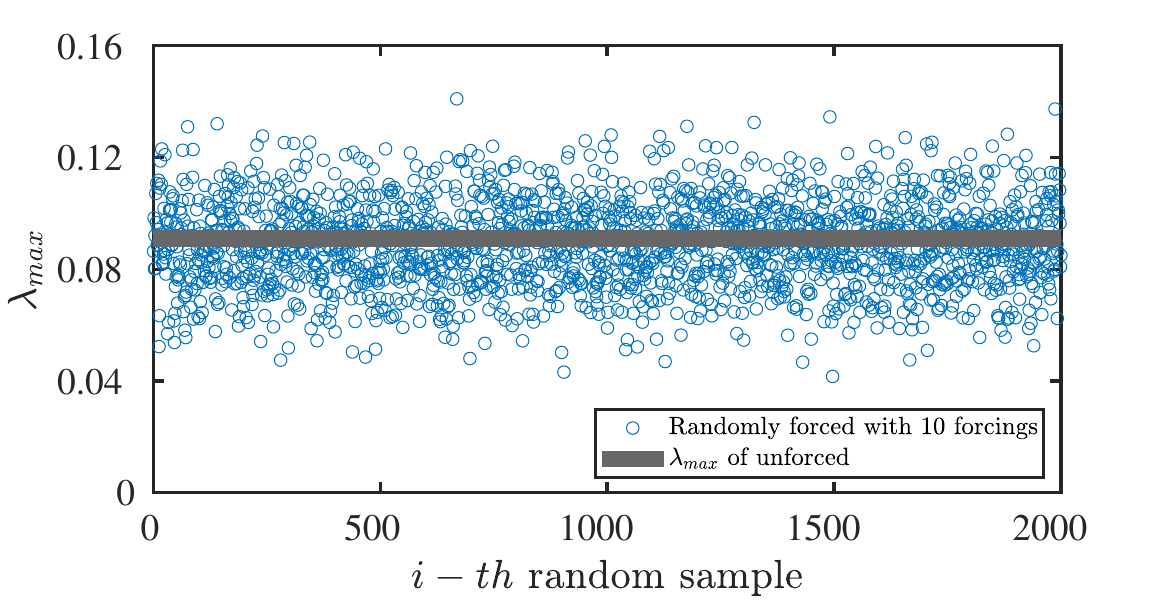}
\caption{ Maximum Lyapunov exponent $\lambda_{max}$ for $2000$ realisations of $m = 10$ stochastic variables.}
\label{fig:lambda_max_10_forcings}
\end{figure}

We now proceed to conduct UQ with se-gPC; the independent stochastic variables are the $m = 10$ amplitudes $\Phi_i$, as already mentioned. We first check if the system maintains its chaotic behaviour with the stochastic forcing. To this end, the system was integrated  for $2000$ random $\Phi_i$ inputs over a time horizon of $T = 200$  and random initial conditions for each input. The Lyapunov exponents were computed for each realisation using the methodology presented in \cite{Lyap_QR}; the  maximum exponent $\lambda_{max}$ is shown in figure \ref{fig:lambda_max_10_forcings}. For all realisations the forced system maintained its chaotic behaviour, with a maximum Lyapunov exponent that fluctuates around the value of the unforced system (solid black line). For all systems it was observed that $\lambda_{max}>0.04$.

\begin{figure}[ht]
\centering
\includegraphics[scale=0.30, trim={1mm 8mm 1mm 1mm}, clip]{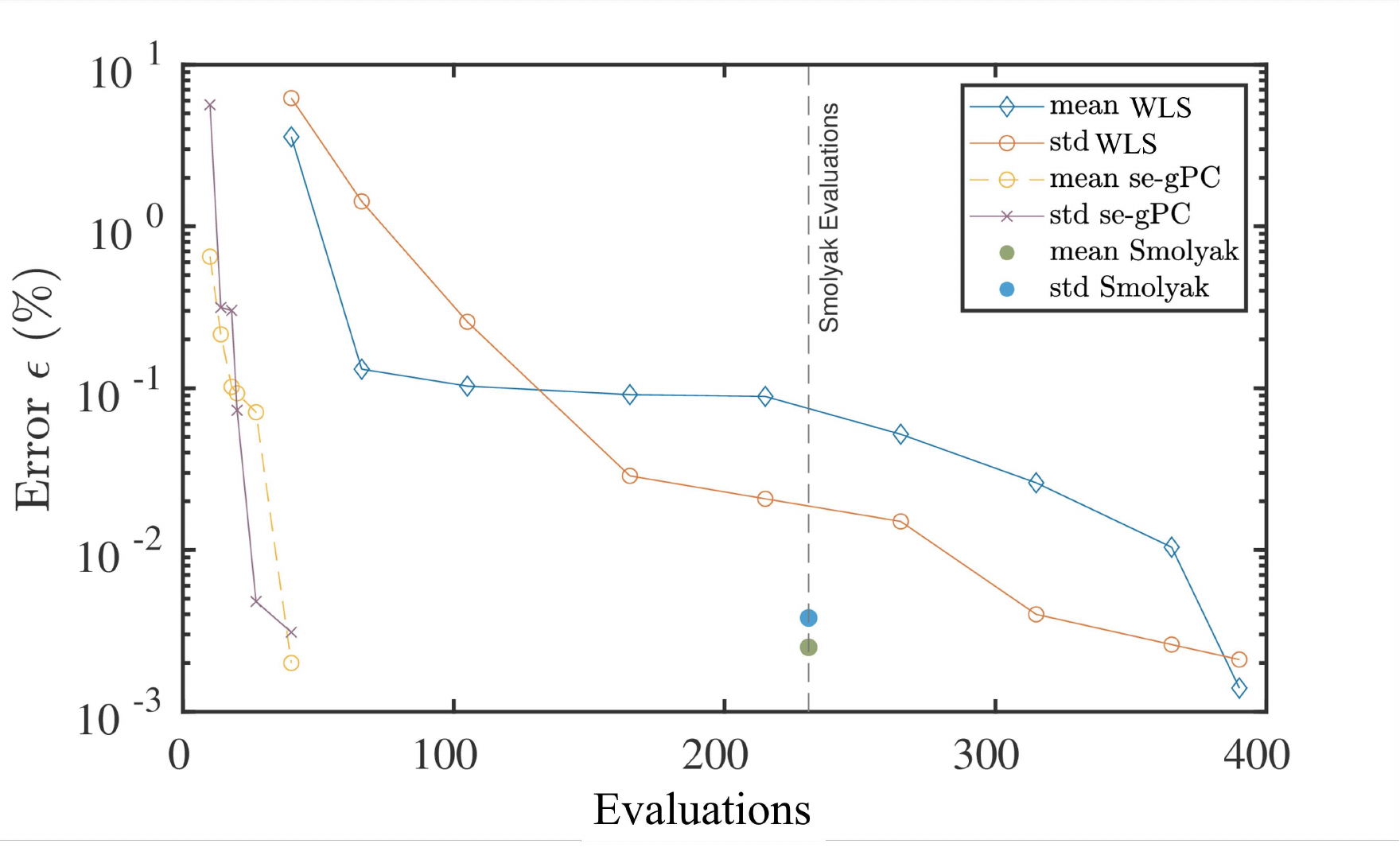}
\caption{Convergence rates of mean and standard deviation of $\overline{J^{(1)}}$ against number of evaluations  computed with WLS, Smolyak Quadrature and se-gPC for $p = 2$. The stochastic input is shown in figure  \ref{fig:KS_forcing_varied}.}
\label{KS_error_convergence_multidim_J1}
\end{figure}

In figure \ref{KS_error_convergence_multidim_J1}, the  convergence rates of the mean and standard deviation of $\overline{J^{(1)}}$ against the number of evaluations required by se-gPC, standard Weighted Least Squares (i.e.\ system \eqref{eq:pce02}) and Smolyak Quadrature are compared. It was assumed that one adjoint solution has the same cost as a direct solution (i.e.\ forward integration of the dynamical system). This is a realistic assumption for the KS system we consider, but generally the cost of obtaining the adjoint solution for a chaotic system is case dependent. In the plot, one adjoint or one forward solution is considered as a single evaluation. The errors are computed with respect to Monte Carlo with $10000$ samples. The plot was obtained with $p=2$,   
where Smolyak Quadrature requires $(m+1)(2m+1)=231$ evaluations.  It is clear that the se-gPC outperforms the other two approaches, providing a very accurate estimation of the mean and the standard deviation with only $40$ evaluations (this corresponds to 20 samples, with $1+m=11$ equations for each sample, in total 220 equations).
  
\begin{figure}[ht]
\centering
\includegraphics[scale=0.30, trim={1mm 8mm 1mm 1mm},
clip]{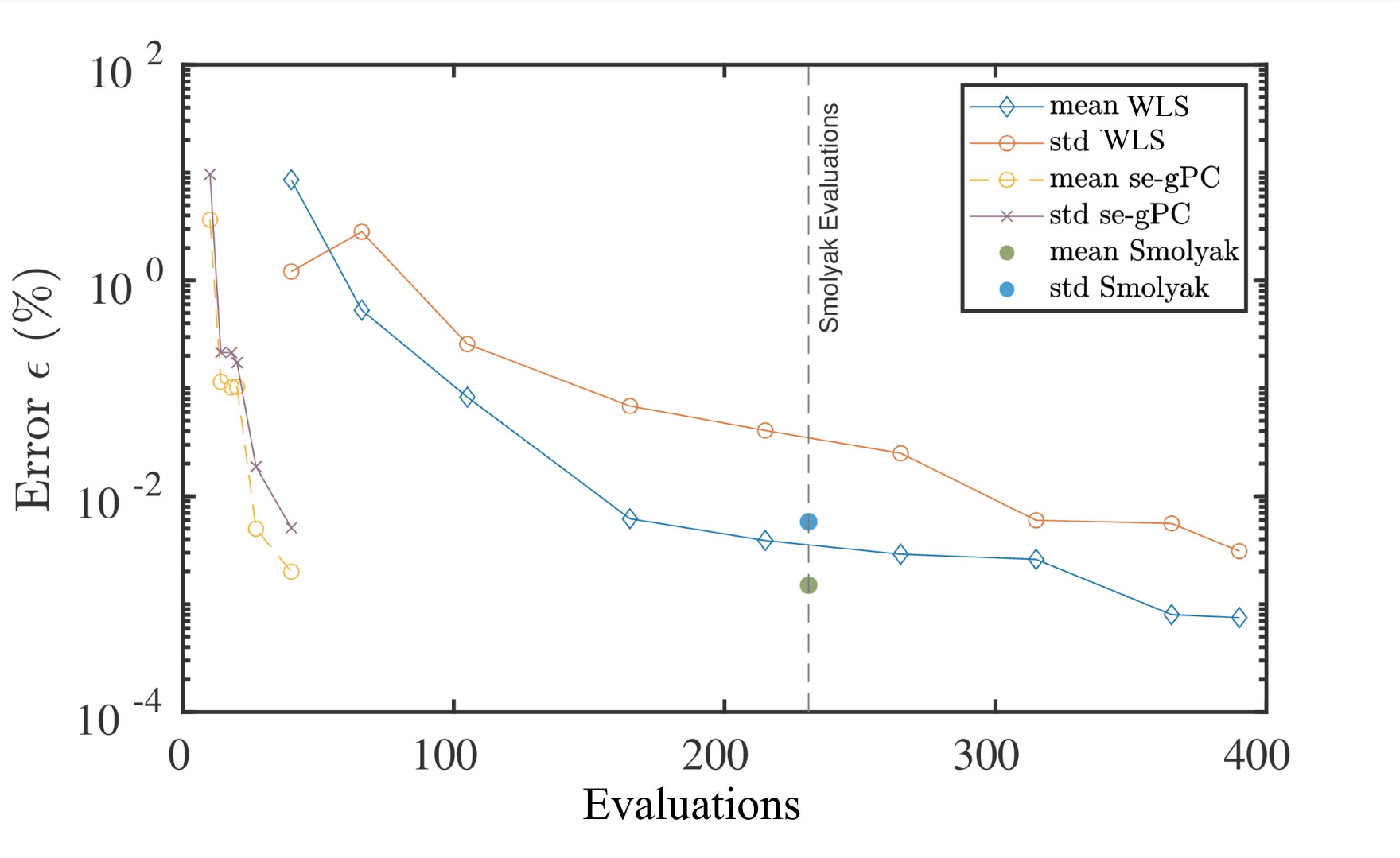}
\caption{Convergence rates of mean and standard deviation of $\overline{J^{(2)}}$ against number of evaluations  computed with WLS, Smolyak Quadrature and se-gPC for $p = 2$. The stochastic input is shown in figure  \ref{fig:KS_forcing_varied}. 
}
\label{KS_error_convergence_multidim_J2}
\end{figure}

The corresponding plot for $\overline{J^{(2)}}$ is shown in figure \ref{KS_error_convergence_multidim_J2}. Once again, se-gPC coupled with the adjoint shadowing operator offers a significant computational advantage compared to other approaches. In practice this allows for the efficient quantification of uncertain input variables on the time-average quantities of chaotic systems. 



\section{\label{sec:6} Conclusions}
We propose a framework for efficient uncertainty quantification of time-average quantities of chaotic systems. We derive the adjoint version of the shadowing harmonic operator for sensitivity analysis of chaotic systems in the frequency domain. We subsequently employ the adjoint to compute the sensitivities of the QoI with respect to all  uncertain variables and use this information to enrich the weighted least squares system from which the spectral coefficients of  polynomial expansion are computed. The adjoint formulation provides all the required sensitivities in a single step, thus significantly increasing the computational efficiency of the method. The sampling points to integrate the dynamical system are obtained by the QR decomposition of an appropriate weighted matrix.  
The computational cost of the method is independent of the number of stochastic variables for polynomial order $p = 1$. 

The adjoint formulation was applied to the Kuramoto-Sivashinsky equation and found to produce accurate sensitivities with respect to the amplitude of  bell-shaped forcing.  When these sensitivities were used to augment gPC, the resulting first and second moments computed matched excellently with Monte Carlo results. We then tested the method on a stochastically forced system with $10$ independent input variables that  determined the actual shape of the forcing. The adjoint was found to produce sensitivities that are in excellent agreement with finite differences and the se-gPC outperformed other UQ methods.  

These attributes make the proposed method a promising candidate for application to chaotic systems with a large number of stochastic inputs.  

\begin{acknowledgments}
K.D.K. acknowledges the financial support of the President’s Scholarship Award from Imperial College London.
\end{acknowledgments}





\bibliography{apssamp,References_GP}

\end{document}